\documentclass[11pt]{article}
\usepackage{stmaryrd}
\usepackage{amssymb}
\usepackage{color, amsmath,amssymb, amsfonts, amstext,amsthm, latexsym}

\usepackage{amssymb, epsfig, amssymb, latexsym}
\usepackage{amsmath}
\usepackage{graphicx}
\usepackage{longtable}
\usepackage{graphicx}
\usepackage{subfigure}
\usepackage{cite}

\usepackage[colorlinks=false, linktocpage=true]{hyperref}

\renewcommand{\phi}{\varphi}

\begin{document}

\title{The maximum likelihood climate change for global warming under the influence of  greenhouse effect and  L\'evy noise}
% \footnote{This work was supported by the  NSFC
%grants  11801192, Leibniz-DAAD research fellowships, 2018(57423756) and Hubei provincial postdoctoral science and technology activity project.} }

\author {Yayun Zheng$^{1, 2, 4}$,   Fang Yang$^{1}$, Jinqiao Duan$^{3*}$, Xu Sun$^{1}$, Ling Fu $^{2}$, J\"urgen Kurths$^{2, 4, 5}$\\
\\$^1$ School of Mathematics and Statistics and Center for Mathematical Science, \\ Huazhong University of Science and Technology, Wuhan, 430074, China.
\\$^2$ Wuhan National Laboratory for Optoelectronics, Huazhong University of \\ Science and Technology, Wuhan, 430074,China.
\\$^3$ Department of Applied Mathematics, Illinois Institute of Technology,\\  Chicago, IL 60616, USA.
\\$^4$ Potsdam Institute for Climate Impact Research, Potsdam, 14473, Germany.
\\ $^5$Department of Physics, Humboldt University, Berlin, 12489, Germany.
\\$^*$Corresponding author. Email: duan@iit.edu
}

\date{}

\maketitle

\begin{abstract}
An abrupt climatic transition  could be triggered by a single extreme event, an $\alpha$-stable non-Gaussian L\'evy noise  is regarded as a   type of noise to generate such extreme events. In contrast  with the classic Gaussian noise, a comprehensive approach of the most probable transition path  for systems under $\alpha$-stable L\'evy noise is still lacking. We develop here a  probabilistic framework, based on  the nonlocal Fokker-Planck equation, to investigate  the maximum likelihood climate change for  an energy balance system under the influence of  greenhouse effect and  L\'evy fluctuations.  We find that a period of the  cold climate state can be interrupted by a sharp shift to the warmer one due to  larger noise jumps, and the climate change for warming $1.5\rm ^oC$ under an enhanced greenhouse effect generates a step-like growth process. These results provide  important insights into  the underlying mechanisms of abrupt climate transitions triggered by a L\'evy process.
\end{abstract}

\section{Introduction}

A rare but  most influential phenomenon in climate change is  a sharp shift from one climate state to another\cite{national2013abrupt}. In particular, the last glacial period experienced rapid, decadal-scale transitions from a stadial cold state to an interstadial warm one, called Dansgaard-Oeschger events \cite{NGRIP, niklas2018do}. One proposed explanation for such events  is that they happened when the Earth system reached a critical tipping point. Tipping points are  associated with bifurcations or induced by noise\cite{ashwin2012tipping, lenton2008tipping}. Meanwhile, there is an alternative view that the abrupt climatic changes could be triggered by a single extreme event, such as heatwaves, droughts or storms. In contrast to the case of Gaussian noise, $\alpha$-stable non-Gaussian noise with heavy tail is regarded as a general   class of noise to generate such extreme events \cite{ditlevsen1999}. We will therefore consider   $\alpha$-stable non-Gaussian L\'evy noise in the following study. Although early-warning signals are detected for an upcoming catastrophic change \cite{dakos2008slowing, scheffer2009early, lenton2011early},  it is extremely difficult to predict  a sudden transition. The identification and characterization of the states along the   path of the dynamics  is the crucial step to explore such  abrupt shift events, where a curve connecting two states in the state space is a transition pathway. Our goal then is to study the maximal likely transition path for a climate change model under $\alpha$-stable  L\'evy fluctuations.

There are several available methods to investigate transitions in stochastic systems with Gaussian noise.  For small-noise-induced transitions,   the Freidlin-Wentzell theory of large deviations is often utilized. The minimizer of the Freidlin-Wentzell action functional   provides  the most probable pathway  and the occurrence rate of the rare events \cite{FW1998, wan2013hybrid}.  For stochastic systems with finite noise intensity, the path integral  provides the expression of the conditional propagator for studying  the most probable transition path \cite{hunt1981path}. This  most probable  transition  path can also be approximated by minimizing the Onsager-Machlup action functional \cite{durr1978om, onsager1953fluctuations}.  Particularly,  in a gradient system,  the most probable path follows the minimum energy, which passes  through the unstable manifold at some saddle points \cite{berkov1998mep}.

  Note that numerous studies mentioned above focused on diffusion processes. These processes are the solutions  of stochastic differential  equations (SDEs)  with  (Gaussian) Brownian motion. However,  the  paleoclimatic records indicate  that  random fluctuations in a rapid transition  have a strong non-Gaussian distribution with a heavy tail and intermittent jumps \cite{ditlevsen1999, dakos2008slowing, corral2019power}, and an $\alpha$-stable L\'evy process is  thought to be an appropriate model for such a non-Gaussian process  \cite{sato1999levy}. Unfortunately,
  it is difficult  to obtain the corresponding action functional from existing research results, when it comes to deal with the transition paths in stochastic systems with non-Gaussian L\'evy fluctuations.  Although, the Onsager-Machlup action functional for stochastic dynamical systems under L\'evy noise is derived recently by one of us \cite{chao2018onsager}, the results are valid only for certain L\'evy fluctuations.
  Therefore, it is desirable to develop a framework for describing the   transition paths to stochastic dynamical systems   under non-Gaussian noise, especially   $\alpha$-stable L\'evy  noise.

%In the present paper, we devise an   approach to determine the maximum likelihood transition path, between two metastable states,  for these non-Gaussian  dynamical systems, and then apply to a climate energy balance model.

  Our approach   uses  the Fokker-Planck equations for non-Gaussian  dynamical systems. These  are deterministic equations describing  how probability density functions propagate and evolve.  The nonlocal or fractional Laplacian operator in these equations is the manifestation of    $\alpha$-stable L\'evy  fluctuations. Recently,  we developed a fast and  accurate numerical algorithm      to simulate  nonlocal Fokker-Planck equations under either absorbing or natural conditions \cite{gao2016fokker}.   Meanwhile,  we derived the Fokker-Planck equations for   Marcus SDEs driven by L\'evy processes      \cite{sun2017marcus}.   We  also used a  nonlocal Zakai equation      to examine the most probable path for systems with $\alpha$-stable L\'evy  systems and   continuous-time observations \cite{gao2016zakai}.  Furthermore,  we  devised most probable phase portraits     to capture certain aspects of  stochastic dynamics \cite{cheng2016mp}, and  applied to examine qualitative changes or bifurcation of equilibrium states under  L\'evy noise \cite{wang2018bifur}.

%We consider the most probable dynamics by examining the maximizer $x_m(t)$ for the probability density function. Inspired by this,  we use a specially conditional  probability density to quantify the transition probability from an initial state to a finial state.  The conditional density is based on Fokker-Planck equations  in SDEs under Gaussian or (non-Gaussian) L\'evy  noise. The corresponding Fokker-Planck equations regarding as a deterministic geometric tool can be viewed as the  attempt in the direction of the maximum likelihood transition.

% Note  that the path mentioned here is not $C^2$ continuous.

In order to determine the maximum likelihood transition path, between two   states,  we have derived  the expression    for the conditional probability density $p(x, t | x_0, 0 ; x_{f}, {t_f})$,   for  sample paths with  initial condition $X(0) = x_0$ and a final condition $X(t_f) = x_{f}$  (i.e.,  sample paths connecting the  two states $x_0$ and $x_f$).    The maximizer $x_m (t)$, at each   time instant $t$,  for the conditional probability density $p(x, t | x_0, 0 ; x_{f}, t_f)$ indicates the maximal likely location of the sample  paths. Taking all  times on $[0, t_f]$, the set of maximizers $x_{m}(t)$ constitute a transition path. It can be referred to as the maximum likelihood transition path between two states $x_0$ and $x_f$. It is only a set of $x_{m}$ which makes the density function $p(x, t | x_0, 0 ; x_{f}, t_f)$  maximum. To illustrate this approach of the maximum likelihood transition path, we will study an energy balance climatic system driven by a discontinuous (with jumps) $\alpha$-stable L\'evy process.  Numerical experiments are  conducted to investigate the impact  of the   non-Gaussianity and   greenhouse factor on the maximum likelihood   transition path between a cold glacial state and a warm interstadial state.

\textbf{The significance of the present work:}    (i) We develop a general probabilistic framework  to quantify the  maximum likelihood transition path, especially for systems with   non-Gaussian L\'evy noise;  (ii) We    verify that an  abrupt transition path in the climatic change energy balance system may be  triggered by  $\alpha$-stable L\'evy noise.  We expect to create the maximum likelihood path  as an efficient research tool, which  quantitatively  describes  how the climate changes, and  explain how the greenhouse effect combined with  L\'evy noise  affect the warming of the Earth.  The better understanding of the underlying mechanisms  is crucial to predict an upcoming abrupt climate change.

%Now we first present  our  approach   to determine and investigate the maximum likelihood   transition path between two metastable states non-Gaussian  dynamical systems.

\section{Results}
\label{Results}

\begin{figure}[t]
\begin{minipage}[b]{ \textwidth}
%\leftline{(a)}
\centerline{\includegraphics[height = 7cm, width = 0.95\textwidth]{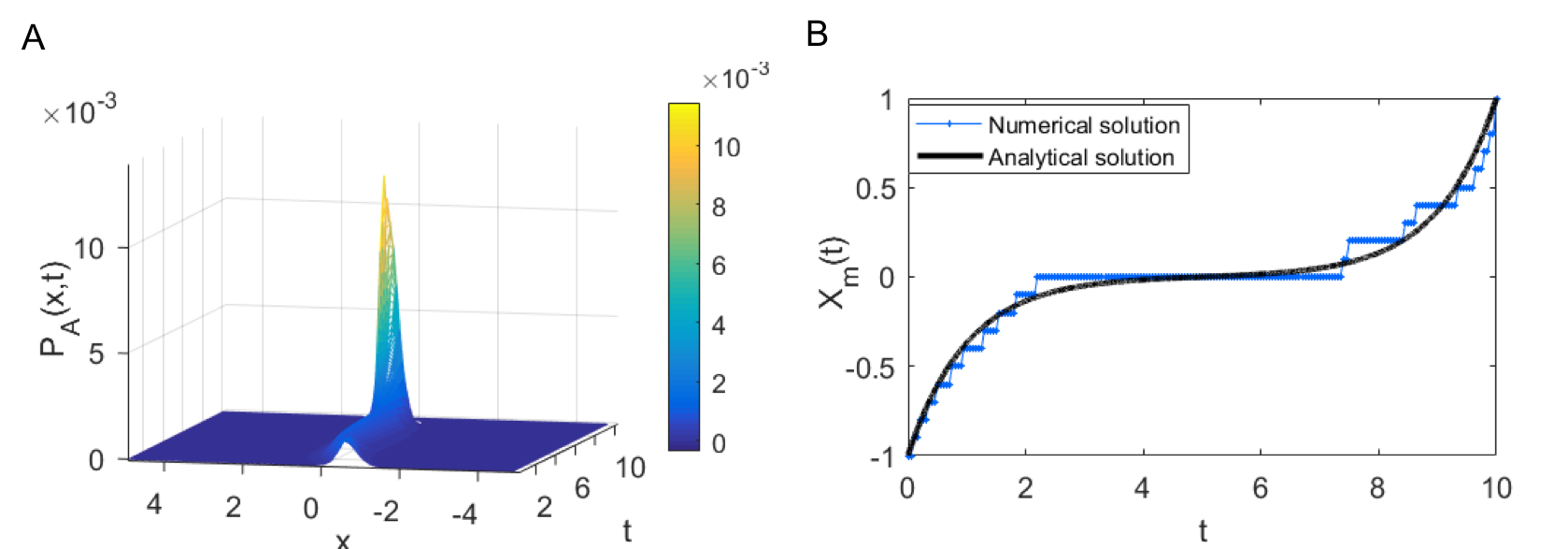}}
\end{minipage}\caption{\textbf{The maximum likelihood transition path.} (A) The conditional probability density function $\mathcal{P}_{A}(x, t)$ for a scalar Ornstein-Uhlenbeck process (see Supplementary Materials S2.).  (B) The numerical simulation of the maximum likelihood transition path is compared with the analytical solution.}
\label{fig:1.2}
\end{figure}

\noindent\textbf {The maximum likelihood transition path.} We  propose  our approach to determine and investigate the  the maximum likelihood  transition path from one state to another  for dynamical systems under  non-Gaussian $\alpha$-stable L\'evy noise. Inspired by   Lemma 3.2 in  \cite{zheng2016delay},   for all $t \in [0, t_f]$ and $x, x_0, x_{f} \in \mathbb{R}^d$, we assume that the conditional probability density function $\mathcal{P}_{A}(x, t)$ given for both the condition $X(0)=x_0$ and $X(t_f)=x_{f}$ exists (where $A$ denotes these two-point conditions).  It can be expressed as
\begin{align}\label{p3}
\mathcal{P}_{A}(x, t) &=  p(X(t)=x | X(0)=x_0; X(t_f)=x_{f})\notag\\
&= \frac{Q(x_{f}, t_f | x, t)Q(x, t | x_0, 0)}{Q(x_{f}, t_f | x_0, 0)},
 \end{align}
where $Q$ is the solution of the associated Fokker-Planck equation with appropriate an initial condition (see Materials and Methods).
The detailed  derivation is given in Supplementary Materials S1.
Subjecting to  the condition $A$, the density function $\mathcal{P}_{A}(x, t)$  has  a  peak at a  time $t \in [0, t_f]$,  and the  peak corresponds to a state $x_{m}(t)$.  It implies that, at a given time instant $t$, the  maximizer $x_m (t)$ for the conditional probability density $\mathcal{P}_{A}(x, t)$ indicates the maximum likelihood location of these stochastic trajectories (or sample paths).  That is, we   find the state $x_m(t)$ by  maximizing the transition  probability density $\mathcal{P}_{A}(x, t)$,
 \begin{equation}\label{p11}
x_{m}(t) = \arg  \max_{x} \mathcal{P}_{A}(x, t).
\end{equation}
Now, we examine the corresponding  transition behavior  by the expression of the conditional  probability density $\mathcal{P}_{A}(x, t)$  in  Eq.(\ref{p3}), in a simple example.  In Fig.\ref{fig:1.2}(A), the conditional probability density $\mathcal{P}_{A}(x, t)$ for a scalar Ornstein-Uhlenbeck process $X(t)$ can be simulated by numerical algorithm of Gao \emph{et al.}\cite{gao2016fokker}  (see Materials and Methods).  Meanwhile, we  examine that the numerical simulation for the  maximum likelihood  path  $x_m(t)$ to be valid, by comparing the numerical solution Eq.(\ref{p11}) (the dashed line)      with the analytical solution (the solid line),  as shown in Fig.\ref{fig:1.2}(B). The calculation of the analytical solution is described in Supplementary Materials S2.

\begin{figure}[t]
\begin{minipage}[b]{ \textwidth}
%\leftline{(a)}
\centerline{\includegraphics[height = 7.5cm, width = 0.9\textwidth]{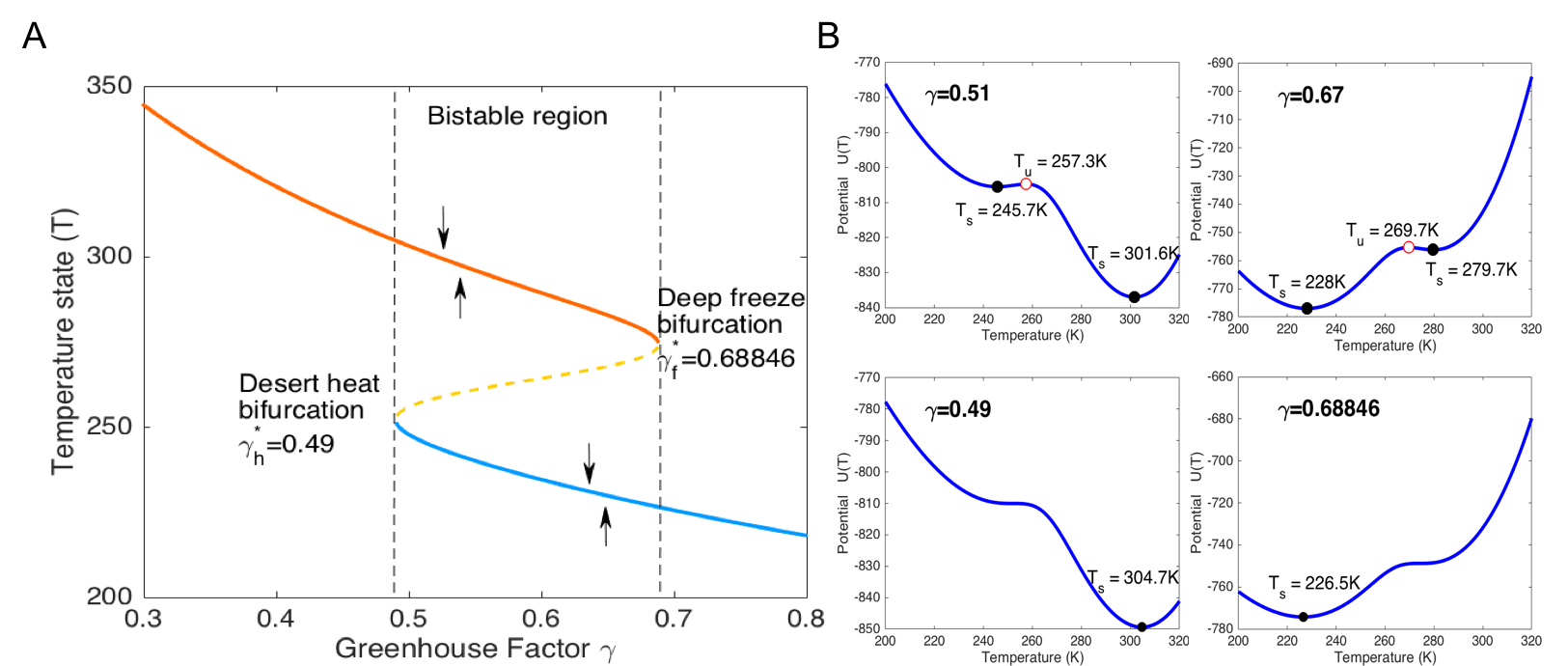}}
\end{minipage}
\caption{\textbf{The energy balance model Eq.(\ref{dbe})  may  have multiple equilibria.} (A) The bifurcation diagram  as the greenhouse factor $\gamma$ varies. (B) The potential functions $U(T)$ for greenhouse factors: $\gamma=0.51$, $\gamma^*_h=0.49$, $\gamma=0.67$, $\gamma^*_f=0.68846$.}
\label{fig:2.1}
\end{figure}

We have thus developed a probabilistic framework  for describing the maximum likelihood transition path between two states. This technique is now applied to study a climate  energy balance model Eq(\ref{dbe}) (see  Materials and Methods) for examining the maximum likelihood climate change. Climate change is represented here by the global mean surface temperature $T(t)$ evolution  throughout the entire system\cite{saltzman2001dynamical, kaper2013matclimat}.

We start the detailed analysis for  the  energy balance model (\ref{dbe}), by examining the equilibrium temperature states: $\rm dT/ \rm dt=0$. The equilibria against the greenhouse effect $\gamma$ shows on the $S$-shaped curve  in Fig.\ref{fig:2.1}(A). The two turning points $\gamma_f^*$ and $\gamma_h^*$ mark the critical parameter values for which branches of equilibria meet and vanish. During the bistable region ($\gamma_h^* < \gamma < \gamma_f^*$),  the deterministic climatic system exhibits the two stable states at $T_s$ and an unstable state at $T_u$ by the potential functions $U(T)$, such as $\gamma = 0.51, 0.67$   shown in Fig.\ref{fig:2.1}(B).
Assuming that  the greenhouse effect    $\gamma$ increases in the bistable region,  the temperature of the cold glacial state and  the warmer interstadial state  are decreasing, which causes the Earth's temperature to drop.  At  $\gamma_f^* \approx 0.68846$,  there is only one stable state $T = 226.5\rm K$ $(-40.65\rm ^oC)$, i.e., the climate will reflect the long time stabilization in an ice-covered state called ``Snowball Earth''  \cite{hoffman1998neoproterozoic}.  In contrast, as the greenhouse factor  $\gamma$ decreases, the temperature of the stable equilibrium states  increase until  the greenhouse effect becomes strong enough at   $\gamma_h^* \approx  0.49$, the Earth will  then remain in  a  high temperature environment $T = 304.7\rm K$ $(31.5\rm ^oC)$.  Thus, the greenhouse factor  values $\gamma_f^*$ and $\gamma_h^*$ are referred to as \textit{deep freeze bifurcation } and \textit{desert heat bifurcation}, respectively \cite{imkeller2001ebm}. These imply that the global surface temperature $T$ increases as the greenhouse factor $\gamma$ decreases. Therefore,  the decreased greenhouse factor enhances the greenhouse effect and  causes a global mean surface temperature rising.

%%. K=46.8\approx

The climate change of underlying extreme events can be modeled by the stochastic energy balance system (\ref{sbe}) driven by a symmetric  $\alpha$-stable  L\'evy process in  Materials and Methods. Here, the $\alpha$-stable  L\'evy process $L^\alpha_t$ is a pure jump process defined by a stable   L\'evy process with $0 < \alpha <2$. The detail introduction for the $\alpha$-stable  L\'evy process is given by Supplementary Materials S3. Next, we discuss the impact of $\alpha$-stable  L\'evy noise on the climate change. The corresponding change behavior of global temperature  is present by numerical simulation of  the maximum likelihood transition path $x_m(t)$ in Eq.(\ref{p11}).

\begin{figure}[t]
\begin{minipage}[b]{ \textwidth}
%\leftline{(a)}
\centerline{\includegraphics[height = 10cm, width = \textwidth]{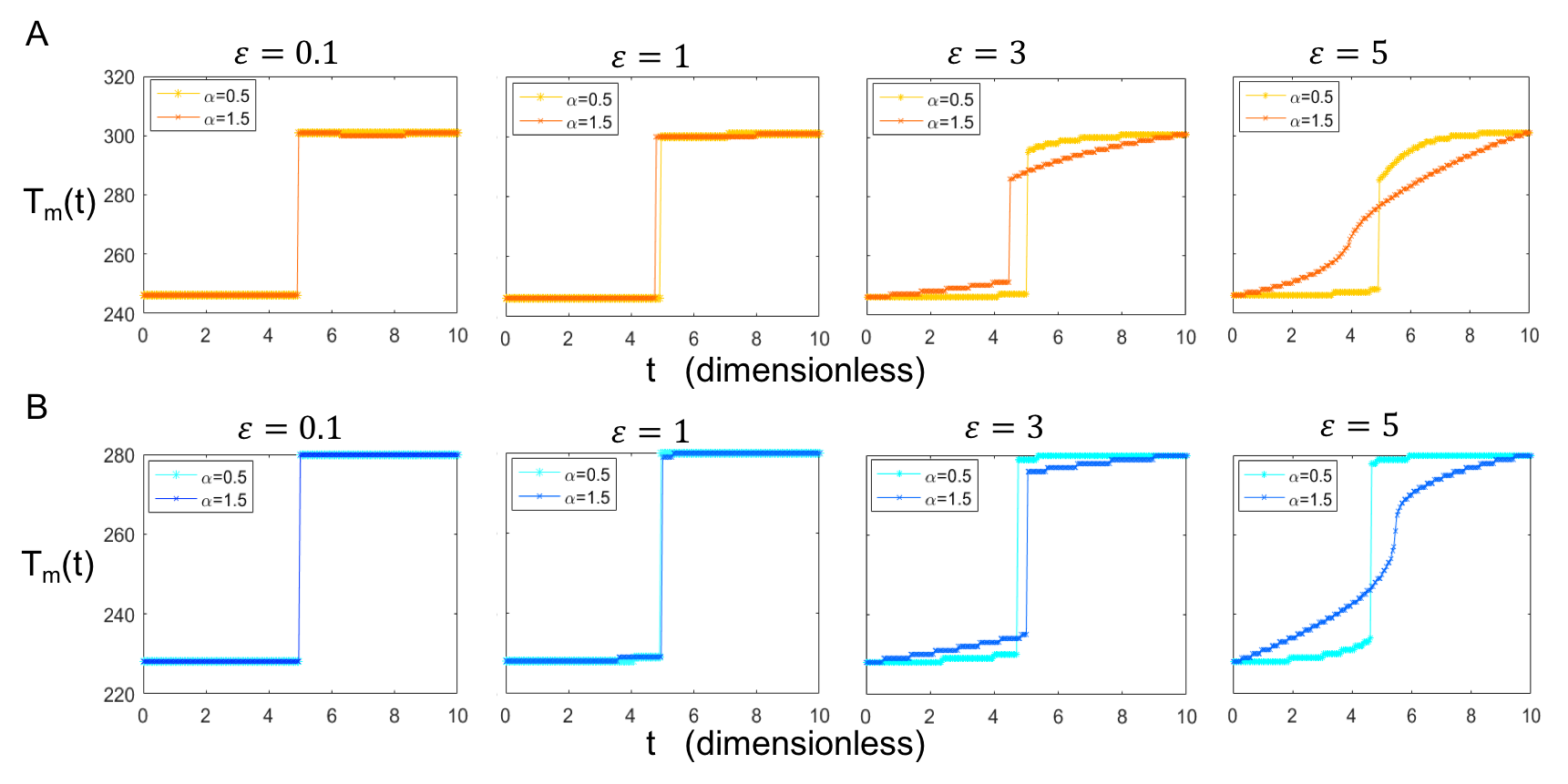}}
\end{minipage}
\caption{\textbf{The dependence of the maximum likelihood climate change on the $\alpha$-stable L\'evy noise intensities $\epsilon$ for global warming of $1.0\rm ^oC$.} The maximum likelihood  path for:  (A) $\gamma=0.51$ (transition from the cold climate stable state $T=245.7\rm K$ to the warmer one $T=301.6\rm K$) and (B) for $\gamma=0.67$ (transition from the deep-frozed climate stable state $T=228\rm K$ and the warmer one $T=279.7\rm K$) with $\alpha=0.5$ and $\alpha=1.5$.}
\label{fig:3.1}
\end{figure}

\medskip
\noindent \textbf{Effect of  $\alpha$-stable L\'evy  noise for global warming of $1\rm ^oC$.}
The heat capacity  $C=46.8\rm Wyrm^{-2}$ is regarded as a weighted average value of ocean and land surface warming by $1.0\rm ^oC$\cite{hansen2005capacity}.  We investigate the pathway of climate change starting in a cold glacial state and landing in a warmer interstadials state, when the climate   system is triggered by a single extreme event. Two kinds of typical greenhouse factors near the bifurcation points are chosen, such as $\gamma=0.51$ and $\gamma=0.67$, each of them has two stable states as shown in Fig.\ref{fig:2.1}(B).  The numerical results of the maximum likelihood  transition path are presented with different amounts of $\alpha$-stable L\'evy  noise intensities $\epsilon = 0.1, 1, 3,  5$ in Fig.\ref{fig:3.1}. We compared the transition path for two representative values of  L\'evy index $\alpha  = 0.5$ and $\alpha = 1.5$, which  corresponds to  larger jumps with lower frequencies and smaller jumps with higher jump probabilities, respectively.

Firstly, we consider the  maximum likelihood transition path from  the cold glacial state ($T = 245.7\rm K$) to the warmer interstadial one ($T = 301.6\rm K$) for $\gamma=0.51$ as shown in Fig.\ref{fig:3.1}(A). We choose the bounded domain $D = (208\rm K, 308\rm K)$ because  the size of the basin of the cold glacial state is equal to the warmer one.  For  small  values of $\epsilon$ ($\epsilon \leqslant 1$),  we find that a period of  a relatively stable  cold glacial state is interrupted by a sharp transition to the warmer interstadial state.  The path of the  maximum likelihood transition is not obviously different between $\alpha=0.5$ and $\alpha=1.5$.  A significant difference is presented as  $\epsilon$ is increasing, such as $\epsilon=3$ and $5$. For $\alpha=0.5$, we find that there is a sudden jump when the global surface temperature  gradually increases from the cold state to  the warmer one.
 However, for  $\alpha=1.5$,   the path of climate change  presents   almost continuous growth curve as   $\epsilon=5$. The results on the  maximum likelihood transition path show that one has to consider both the value of $\epsilon$ and  $\alpha$ when deciding which of the three competing factors plays a vital role in the climate change system, the noise intensity, the jump frequency, or the  jump size.

Next, Fig.\ref{fig:3.1}(B) illustrates the behavior of the maximum likelihood transition for $\gamma=0.67$ from  the ice-cover state ($T = 228\rm K$) to the warmer interstadials one ($T = 279.7\rm K$). Given a bounded domain $D = (220\rm K, 320\rm K)$ with the same size of the attraction basin of the cold and warmer state,  we uncover that the climate likely experiences a rapid increase  followed by a long-time stable trend  for  small noise intensities. This change behavior is in agreement with the corresponding result for $\gamma=0.51$.  However,  for $\gamma=0.67$, an obvious difference is that the   transferring time  to the warmer state  is longer than  for $\gamma=0.51$ in  the $\epsilon=5$ and $\alpha=1.5$ case. The reason is that the potential  of the cold state is larger than  the warmer one  for $\gamma=0.51$, thus the climate change system is easier to shift from higher  to  lower.  For both $\gamma=0.51$ and $\gamma=0.67$,  the   transition paths   present the characteristic of  a sharp shift   when  the L\'evy index $\alpha = 0.5$.   This implies that $\alpha$-stable L\'evy  noise  is easier to induce an abrupt transition in the case of larger jumps with lower jump frequencies. Note that  the critical shift point is reached nearing $t = t_f / 2$ for $\alpha=0.5$.  It can be proved by the expression of the conditional probability density $\mathcal{P}_{A}(x, t)$ for  $\alpha$-stable  L\'evy process in  Eq.(\ref{p3}). For $0 < \alpha <1$, we can prove that  the maximal value of $\mathcal{P}_{A}(x, t)$ is reached at states $x_0$ and $x_f$ when  the time instant $t = t_f / 2$, simultaneously.

\begin{figure}[]
\begin{minipage}[b]{ \textwidth}
%\leftline{(a)}
\centerline{\includegraphics[height = 13cm, width = 0.95\textwidth]{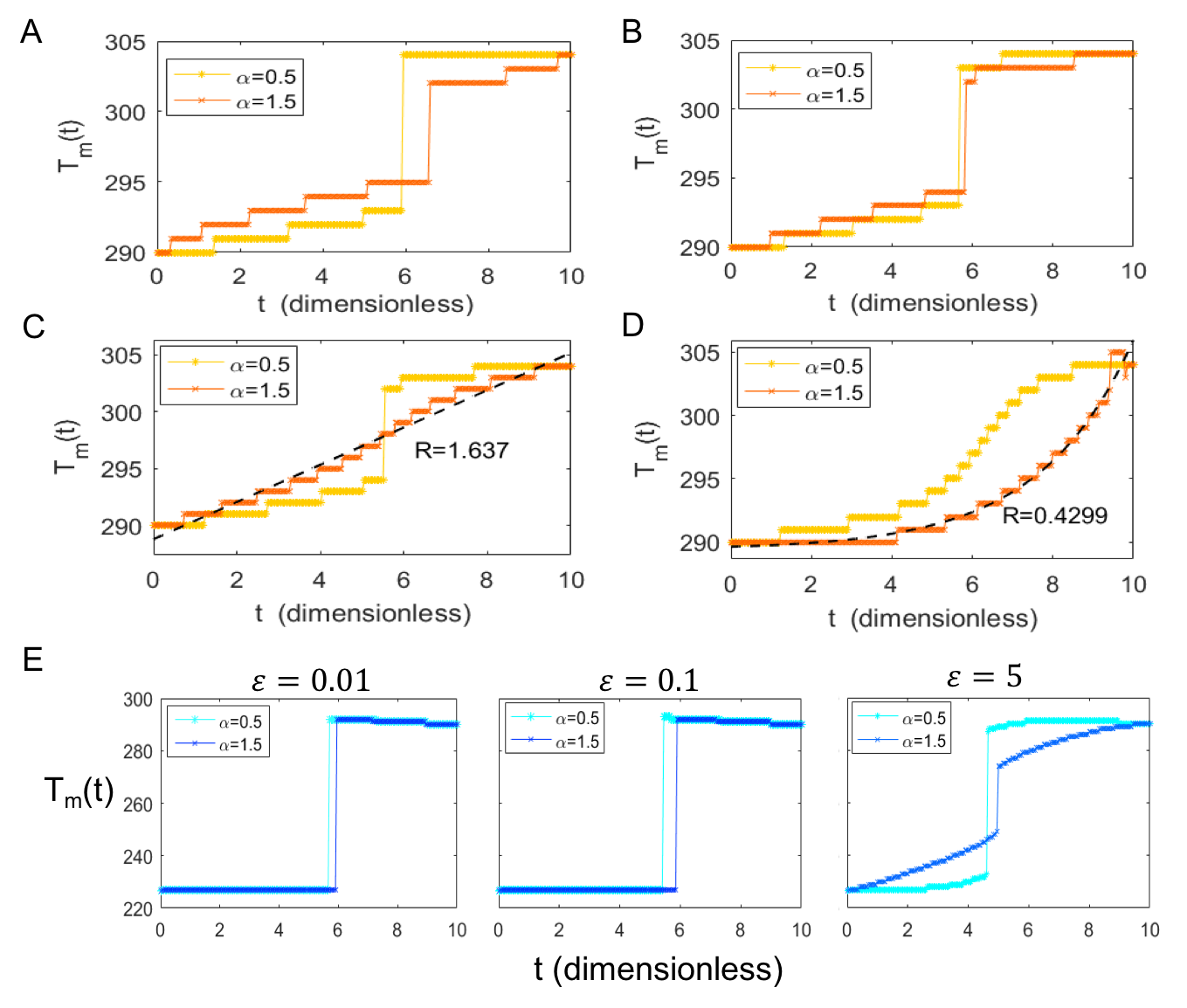}}
\end{minipage}
\caption{\textbf{The effect of  $\alpha$-stable L\'evy noise  intensities on  the maximum likelihood transition path for bifurcation greenhouse factor.} Transition from the current state $290\rm K$ to the stable state $304.7\rm K$ of desert heat bifurcation $\gamma_h^*=0.49$ with  (A) $\epsilon=0.01$. (B) $\epsilon=0.1$. (C) $\epsilon=1$. (D) $\epsilon=5$. (E) Transition from the stable state  $226.5\rm K$ of  deep-frozed bifurcation $\gamma_f^*=0.68846$ to the current state $290\rm K$.}
\label{fig:4.1}
\end{figure}

\medskip
\noindent \textbf{Effect of  $\alpha$-stable L\'evy  noise for global warming of $1.5\rm ^oC$.}
Due to human activities,  the world has already warmed by $1.0\rm ^oC$ since the pre-industrial times. The Intergovernmental Panel on Climate Change(IPCC) special report on the impact of global warming of $1.5\rm ^oC$  has caught   broad attention. According to the report, the global warming is likely to reach $1.5\rm ^oC$ between 2030 and 2052 year if we continue to increase greenhouse gas emission at the current rate\cite{IPCC}.

Meanwhile, we look at the effect of warming  $1.5\rm ^oC$ on the climate most likely changes by changing the heat capacity to $C=70.27\rm  Wyrm^{-2}$, while keeping other factors the same, i.e., $\gamma=0.51$, $0.67$ with varying  $\epsilon = 0.1, 1, 3, 5$ and  $\alpha=0.5, 1.5$. From the numerical results shown in Fig.S1(see Supplementary Materials),
we reveal that the   maximum likelihood  transition paths for global warming  $1.5\rm ^oC$ have the  similar behavior to the case of  $1.0\rm ^oC$ as mentioned in the previous section (Fig.\ref{fig:3.1}).  Clearly, the climate dynamics  subjecting to a small  noise intensity, such as $\epsilon = 0.1$, or a small value of $\alpha$, such as $\alpha=0.5$,  experience occasional sharp transition from one state to another. Meanwhile, the transition path presents a nearly continuous trend attributing to small jumps combined with larger noise intensity, such as $\alpha=1.5$ and $\epsilon=5$.  On the other hand, there are some subtle differences when the global warms $1.5\rm ^oC$, such as $\gamma=0.51$ with $\epsilon=3$ and $\alpha=0.5$. In contrast to the temperature continuing to increase in  the case of warming $1.0\rm ^oC$ (Fig.3(a)), the temperature for warming $1.5\rm ^oC$ has a performance of rapid decrease following by slow increase after a sudden shift.  This implies that it is more likely to induce the multiple abrupt climate changes when the global surface temperature warms to  $1.5\rm ^oC$.

Comparing the maximum likelihood transition paths for $\gamma=0.51$ and $\gamma=0.67$, we find that the growth of temperature for $\gamma=0.67$ tends to the slower, before or after an abrupt transition with varying  parameters $\epsilon$ and $\alpha$. It means that  the weakened greenhouse effect ($\gamma=0.67$) may be  slowing down the climate change. That is the reason why we need to reduce and even cut completely emissions of greenhouse gas. The tendency of slowing warming may help people  to gain   time to adapt to  extreme climate, such as heatwaves, droughts and flooding.

 Furthermore, we focus on  how the  climate changes  from the current temperature state to the high-temperature one for  global warming $1.5\rm ^oC$. To illustrate such question, we examine the  effect of  $\alpha$-stable  L\'evy noise  on  the maximum likelihood  transition path from a current state $T= 290\rm K$ (global average temperature in April 2019) to a warmer  state. The warmer  state is considered as the stable state $T= 304.7\rm K$ of the desert heat bifurcation $\gamma_h^*=0.49$,  which corresponds to the enhanced greenhouse effect.

 \begin{figure}[!t]
\begin{minipage}[b]{ \textwidth}
%\leftline{(a)}
\centerline{\includegraphics[height = 10cm, width = 0.95\textwidth]{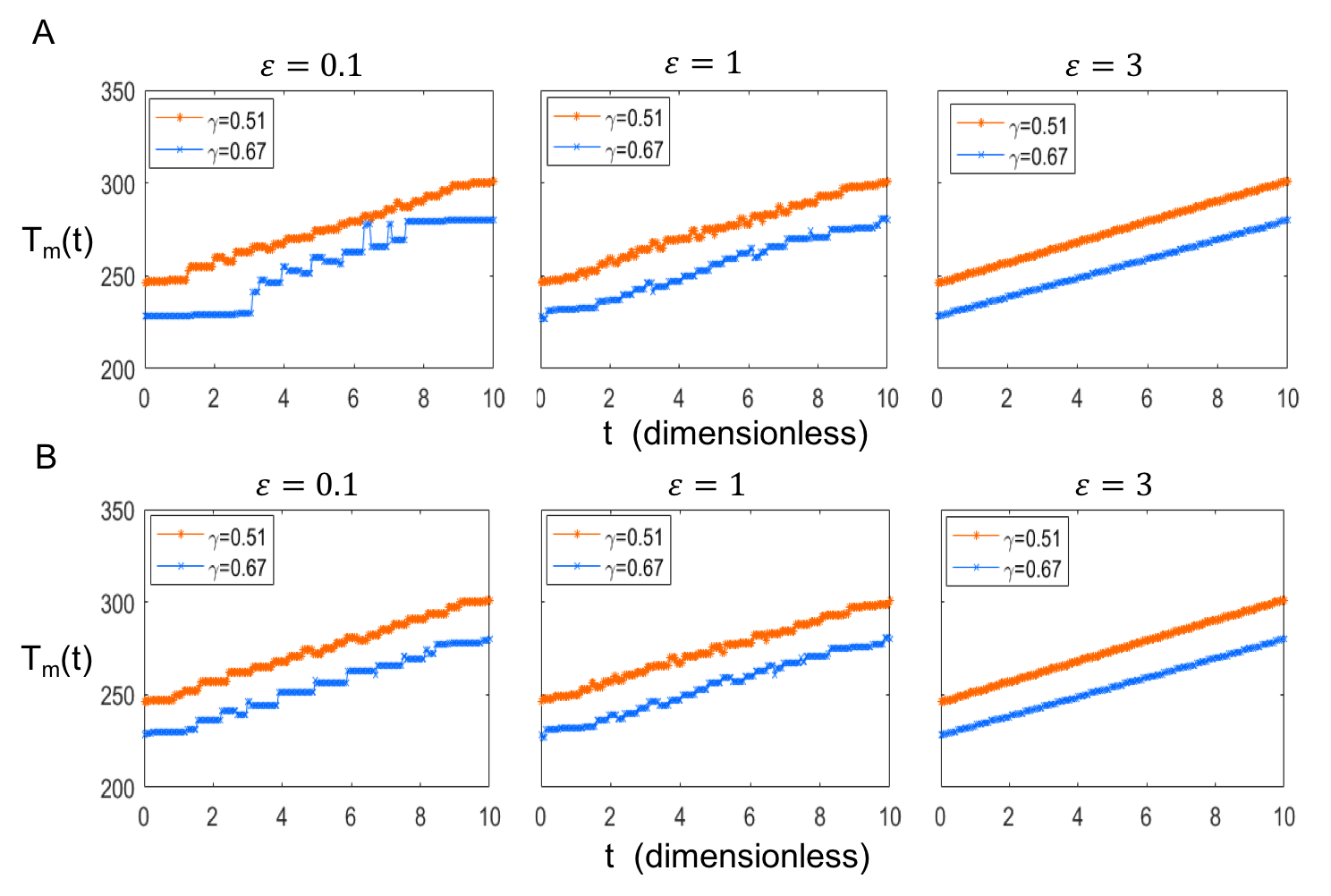}}
\end{minipage}
\caption{\textbf{The effect of  $\alpha$-stable L\'evy noise  intensities on  the maximum likelihood climate change for deep-frozed bifurcation $\gamma_f^*=0.68846$} . Transition from the stable state  $226.5\rm K$ of $\gamma_f^*$ to the current state $290\rm K$.}
\label{fig:5.1}
\end{figure}

In Fig.\ref{fig:4.1}, we find that the maximum likelihood transition path is a step-like process increasing  to the warmer  state from the current one.
For  $\alpha=0.5$, the global surface temperature  manifests itself as a stepwise slowly increasing followed by an abrupt shift, and the magnitude of the suddenly jump gradually decreases as the $\epsilon$ increases. In contrast, for a larger  $\alpha$, such as $\alpha=1.5$, the temperature is increasing nearly linearly for $\epsilon=1$  with the transfer rate $R=1.637$ as shown in Fig.\ref{fig:4.1}(C). We give an expression of the linear growth by curve fitting with $0.95$ confidence bounds,
  \begin{equation*}
T_m(t) = 1.637t+288.8.
\end{equation*}
Figure.\ref{fig:4.1}(D) shows that  the climate change is not a simple linear relationship with  time $t$ when the noise intensity $\epsilon$ increases to $5$.  The temperature maintains around the current state at the outset, then the growth rate changes slowly and then stabilizes at the warmer state $T=304.7K$. In view of such characteristics, an exponential growth function is proposed to fit the climate change,
\begin{equation*}
T_m(t) = 0.2252e^{0.4299t}+289.4.
\end{equation*}

 In the following discussion, we pay attention to  the maximum likelihood pathways of climate change  from a  frozen state to the current state, due to global warming $1.5\rm ^oC$, the collapse of ice sheets leads to  rising of sea  level.  In Fig.\ref{fig:4.1}(E), we look at the effect of  $\alpha$-stable L\'evy  noise on  the maximum likelihood path transition from  the stable state $T=226.5\rm K$ of the deep-frozed bifurcation $\gamma_f^*=0.68846$ to the current one $T=290\rm K$.  In comparison with ``linear''  growth of temperature for  the greenhouse factor $\gamma=0.51$, we find that the maximum likelihood  path for   the greenhouse factor $\gamma=0.67$ always presents a sudden transition between two  states even for larger noise intensity $\epsilon$. The global surface temperature  maintains in the frozen state for a period of time, then it   decline slowly to the current temperature after a sudden jump.

%%%. K=70.27

%%Brownian motion

\medskip
\noindent \textbf{Effect of the Gaussian noise.} Having discussed the maximum likelihood transition path under  $\alpha$-stable L\'evy noise, we now examine the effect of  Gaussian noise on the  maximum likelihood pathway. Figure.\ref{fig:5.1} shows that the maximum likelihood  path has a similar transition behavior for global warming  $1.0^oC$ and $1.5^oC$ respectively,  with varying Gaussian noise intensity $\sigma$.   In contrast to sharp transition under non-Gaussian $\alpha$-stable L\'evy noise,
we find that the global surface temperatures under Gaussian noise are gradually increasing along with small fluctuations. Moreover, the temperature increases steadily in a linear  fashion in the case of  noise intensity $\sigma=3$.  For small value of $\sigma$, such as $\sigma=0.1$, the fluctuation amplitude at the greenhouse factor  $\gamma=0.67$ is larger than that  at the greenhouse factor  $\gamma=0.51$.

\section{Discussion}
To understand the mechanism of an abrupt transition in climate change, we have proposed an approach,  based on transition probability densities and nonlocal Fokker-Planck equations,  to investigate the maximum likelihood transition path from a stadial cold state to an interstadial warm one under $\alpha$-stable L\'evy noise.  The maximum likelihood transition path $x_m(t)$ is defined as the maximizer of  the conditional probability density function $\mathcal{P}_{A}(x, t)$, for each time instant $t$,  subject to an initial condition $X(0)=x_0$ and a final condition  $X(t_f)=x_{f}$.

\textbf{Our approach has the following advantages over the existing methods for examining the most probable transition paths:} (i) Our approach expresses the probability density for transition sample paths via solutions of the associated Fokker-Planck equation, and thus avoids the difficulty for obtaining the action functional in the Onsager-Machlup approach,   in  systems with pure $\alpha$-stable L\'evy noise  \cite{durr1978om, chao2018onsager};  (ii) Our approach applies to systems with either Gaussian or non-Gaussian noise, is not an asymptotic method, and thus we avoids the assumption of  sufficiently small noise intensity (which is required in the large deviation approach \cite{FW1998});  (iii)  Our approach applies while a path integral representation for systems with non-Gaussian L\'evy noise is not yet  generally available, as noted in our earlier work \cite{huang2019characterization}.

Applying our  approach to a climate  energy balance system under interaction of greenhouse effect and  $\alpha$-stable L\'evy noise, we examine the  maximum likelihood climate change for global warming of $1.0\rm ^oC$ and $1.5\rm ^oC$, respectively. Numerical simulations have revealed  the delicate  dependence  of the climate change     on the  the noise intensity, the jump frequency and the  jump size for  various   greenhouse factors.   We find that a period of the relatively stable  climate has been interrupted by sharp transitions to the warmer state   attributing to larger jumps with lower frequency. Such a phenomenon implies that \textbf{the discontinuous  jumps of  $\alpha$-stable  L\'evy process  may be thought as  the underlying mechanism  leading to an abrupt shift.} Meanwhile, comparing with two typical greenhouse factors  nearing bifurcation points, we discover  that the weakened greenhouse effect ($\gamma=0.67$) is more effective on slowing down the climate change.

The greenhouse gas emissions related to the global warming $1.5\rm ^oC$  have significant  influences on humanity and ecosystems.
Furthermore, we uncover that  the maximum likelihood  path for an enhanced greenhouse effect generates a step-like growth process, as transferring from the current temperature state to the high-temperature one.  Moreover, we find that the global surface temperature stepwise increases with an exponential transfer rate for larger noise intensity combined with small jumps. However, for the weakened greenhouse effect, the climate  suddenly reaches the warmer interstadials state in which it has been for a long time at  the frozen state.

Finally,  as a comparison, we have also examined the maximum likelihood climate change when the energy balance system is under  Gaussian  fluctuations. The global surface temperature are gradually increasing by changing in the small fluctuations. The  continuous  sample paths  could explain  why  the climate models  with Gaussian fluctuations can not succeed in describing  sudden shifts between climatic states.

\section{Materials and Methods}

\medskip
\noindent\textbf{Fokker-Planck equation for  the transition probability density  $ Q(x ; t |\xi , s)$.}
In this section, we introduce the Fokker-Planck equation for  the transition probability density.
For the sake of explanation, we consider the following  SDE  with the constant noise intensity $\eta >0$
\begin{equation}\label{p0}
{\rm d}X(t) = f(X(t)){\rm d}t+ \eta {\rm d}N(t) , \quad X(0)=x_0 \in \mathbb{R}^d,
\end{equation}
where $X(t) = (X_1(t), X_2(t), \cdots, X_d(t))$ is a $\mathbb{R}^d$-valued stochastic process, and $f : \mathbb{R}^d \rightarrow \mathbb{R}^d$ is Lipschitz continuous. The $\mathbb{R}^d$-valued noise processes $N(t)$ is either a   standard Brownian motion $B(t)$, or a symmetric $\alpha$-stable L\'{e}vy process  with L\'{e}vy index $\alpha \in (0, 2)$.
Two types of probability density expressions are introduced:  $p(X(t)=u)$ is a general case to represent the density for the $\mathbb{R}^d $-valued solution $X(t)$ of SDE at $X(t)=u$; and $Q(u ; t | \xi , s)$ is reserved to denote the transition density, which is defined as $Q(\cdot , \cdot | \cdot , \cdot):  \mathbb{R}^d \times [0, t_f] \times \mathbb{R}^d \times [0, t_f] \rightarrow [0, \infty)$.    For example, $Q(u ; t | \xi ; s)$ with $0 \leqslant s < t \leqslant T_f$ represents the density of $X(t)$ at $X(t) = u$ given $X(s) = \xi$. It can be expressed in terms of $p$,
\begin{equation*}
Q(u ; t | \xi ; s) = p(x(t)=u | x(s)= \xi).
\end{equation*}

We suppose that  for each $x_0 \in \mathbb{R}^d$, the SDE  (\ref{p0})  has  a unique strong solution, and the probability density for this solution exists and is strictly positive. Then the transition probability density $Q(x ; t | \xi , s)$ for   (\ref{p0}) with Brownian motion ($N(t)=B(t)$) satisfies a Fokker-Planck equation \cite{klebaner2012introduction}
\begin{align}\label{p1}
%\begin{cases}
   \frac{\partial }{\partial t} Q(x ; t |\xi , s) & = - \sum^{d}_{i=1} \frac{\partial}{\partial x_i}(f_i(x)Q(x ; t | \xi , s))  + \frac{\eta^2}{2} \sum^{d}_{i, j=1} \frac{\partial^2}{\partial x_i x_j}Q(x ; t | \xi , s).
     % \end{cases}
\end{align}
The transition probability density  for  (\ref{p0}) driven by an $\alpha$-stable L\'evy motion ($N(t)=L^\alpha_t$)   satisfies  the following nonlocal Fokker-Planck equation \cite{duan2015introduction}
\begin{align}\label{p2}
%\begin{cases}
  \frac{\partial }{\partial t} Q(x ; t | \xi , s) & = - \sum^{d}_{i=1} \frac{\partial}{\partial x_i}(f_i (x)Q(x ; t |\xi , s)) \notag\\
 &+ \eta^\alpha \int_{\mathbb{R}^{d}\backslash \{0\}}[Q(x +y; t | \xi , s) - Q(x ; t | \xi , s) - \sum^{d}_{i=1} \frac{\partial}{\partial x_i}I_{\|y\|<1} y  Q(x ; t | \xi , s)] \nu_\alpha(dy) .
     % \end{cases}
\end{align}
The integral part of the  right hand side  is actually  the nonlocal or fractional Laplacian operator, reflecting the  non-Gaussian  $\alpha$-stable L\'evy  fluctuations \cite{duan2015introduction}.
Both of them fulfill the same initial condition
\begin{equation*}
\lim_{t \rightarrow s} Q(u , t | \xi , s) = \delta (u - \xi).
\end{equation*}

\medskip

\noindent\textbf{Conditional probability density $\mathcal{P}_{A}(x, t)$ for sample paths connecting two states.}
The  conditional probability density $\mathcal{P}_{A}(x, t)$ is defined in a $(x, t, p)$-space , and is used to describe the probability density of the SDE  (\ref{p0}), i.e., $X_t$ is located at the position $x$ at time $t$ subjecting to condition $A$.
 Here,  $x$  is a state in the full phase space in which the dynamics is Markovian, the subscript $A$ is  used to indicate the special type of constraint with the initial condition $X(0) = x_0$ and  the  final condition $X(t_f) = x_{f}$.
We can now construct the conditional probability distributions based on the expression of transition probability density $Q$ in term of the solutions of the associated Fokker-Planck equation (\ref{p1}) or (\ref{p2}).
\begin{equation*}
\mathcal{P}_{A}(x, t) = \frac{Q(x_{f}, t_f | x, t)Q(x, t | x_0, 0)}{Q(x_{f}, t_f | x_0, 0)}.
 \end{equation*}
The detailed derivation of this crucial formula is given in Supplementary Materials S1.

\medskip

\noindent\textbf{Simulation for the maximum likelihood  transition path.} The conditional probability density $\mathcal{P}_{A}(x, t)$  is  related to the  transition density $Q(x_{f}, t_f | x, t)$ of reaching the target state $x_{f}$ and the transition density $Q(x, t | x_0, 0)$ of starting the initial state $x_0$.
Therefore, the numerical calculation for $\mathcal{P}_{A}(x, t)$ can be converted into the calculation of the product of $Q(x_{f}, t_f | x, t)$ and $Q(x, t | x_0, 0)$. Each of them  satisfies the Fokker-Planck equations (\ref{p1}) or (\ref{p2}),  the main simulation problem is how to find a solution in a nonlocal Fokker-Planck equation(\ref{p2}). In the present paper, we apply the ``punched-hole" trapezoidal numerical algorithm of Gao \emph{et al.}\cite{gao2016fokker} to solve the fractional operator  under the absorbing condition. The absorbing condition implies that the density will vanish once it is out of a bounded domain $D$. The probability density $Q(x, t)$ that $X_t$ is located at the position $x$ at time $t$ given the probability profile of its initial position is  Gaussian  $p(x,0) = \sqrt{\frac{40}{\pi}e^{-40x^2}}$.
Finally, the maximum likelihood states $X_m(t)$ can be found via numerical   optimization of $\mathcal{P}_{A}(x, t)$.

It is worth pointing out  how to determine the arrival time $t_f$.  For different random sample trajectories starting at  $x_0$, the time when the system reached the  state $x_f$ is  different. The optimization problem related to the most probable transition path and time $t_f$ are employed by the  theory of  large deviations \cite{wan2017dynamic}.  Here, the arrival time $t_f$ can be determined by Monte Carlo simulations, which  calculate  the average time of arriving at state $x_f$, or by the first mean exit time starting at state $x_0$ from an  interval $D$ as in our earlier work  \cite{zheng2016genetic}.  We emphasize here that the time $t \in [0, 10]$ is dimensionless.

\medskip
\noindent \textbf{Energy Balance  Model.}
 The global energy change is expressed by the difference between the incoming solar radiative energy and the outgoing radiative energy  at time $t$,
\begin{equation}\label{dbe}
C\frac{\rm dT}{\rm dt} = \frac{1}{4}(1 - \alpha(T))S_0 - \gamma \theta T^4.
\end{equation}
The Eq.(\ref{dbe}) can be written as $\dot{T} = -U'(T)$ with the potential function
\begin{equation*}
U(T) = (-\frac{1}{4}S_0(0.5T+2\ln(\cosh(\frac{T-265}{10}))) + \frac{1}{5}\gamma \theta T^5)/C.
\end{equation*}
Here the solar constant $S_0 = 1368 \rm Wm^{-2}$ and  the Stefan constant $\theta = 5.67 \times 10^{-8} \rm W m^{-2} K^{-4}$ from  the Stefan-Boltzmann law.  The heat capacity  $C$ defines as the amount of heat that must be added to the object in order to raise its temperature. The planetary albedo $\alpha(T)$  on temperature is expressed as \cite{kaper2013matclimat}
\begin{equation*}
\alpha(T)=0.5-0.2\tanh(\frac{T-265}{10}).
\end{equation*}
Meanwhile, the greenhouse effect is a natural process that warms Earth's surface. The absorbed energy by greenhouse gases cause the global average temperature to rise\cite{nordhaus1992optimal, hansen2005capacity}.  To maintain an  energy balance, the greenhouse factor $\gamma \in [0, 1]$ is used to express the outgoing energy  reduction.

\medskip
\noindent \textbf{Stochastic Energy Balance  Model.}
Hasselmann's \cite{hasselmann1976stochastic }  idea is that the short-timescale  fluctuating processes, such as wind above the ocean surface, modeled as stochastic processes can be thought of as driving long-term climate variations. From analyzing paleoclimatic data, Ditlevsen \cite{ditlevsen1999, ditlevsen1999anomalous} shows that such fast time-scale noise  contains a component with an $\alpha$-stable distribution. Extreme events, such as heatwaves, droughts, storms as triggering mechanisms for climatic changes can be represented by $\alpha$-stable L\'evy noise\cite{ditlevsen1999}. Therefore, a type of more realistic  energy balance model with underlying extreme events can be written as
 \begin{equation}\label{sbe}
\frac{\rm dT}{\rm dt} = \frac{1}{C}(\frac{1}{4}(1 - \alpha(T))S_0 - \gamma \theta T^4 )+   \frac{\tilde{\epsilon}}{C}{\dot{L}_t^{\alpha}}.
\end{equation}
Here $\dot{L}_t^{\alpha}$ is a L\'evy noise which can be modeled by a scalar symmetric  $\alpha$-stable L\'evy process with the generating triplet $(0, 0, \nu_\alpha)$ i.e., a pure jump motion with $0 < \alpha <2$ (see Supplementary Materials S3).
On the other hand,  the ``normal'' atmospheric fluctuations effected on the energy balance system is modeled by Gaussian noise $\dot{B}_t$,
\begin{equation}\label{sbebm}
\frac{\rm dT}{\rm dt} = \frac{1}{C}(\frac{1}{4}(1 - \alpha(T))S_0 - \gamma \theta T^4 )+  \frac{\tilde{\sigma}}{C} {\dot{B}_t},
\end{equation}
 where  $\tilde{\epsilon} / C = \epsilon$, and $\tilde{\sigma} / C = \sigma$  are the noise intensities of $\alpha$-stable L\'evy process and Brownian motion, respectively.

\section{Supplementary Materials}
\noindent  S1. The  derivation of  the conditional probability density in Eq.(\ref{p3}).

\noindent  S2. A simple example with  analytical solution for the maximum likelihood  transition path.

\noindent  S3. The $\alpha$-stable  L\'evy process.

\noindent  Fig. S1. The dependence of the maximum likelihood climate change on the $\alpha$-stable L\'evy noise intensities $\epsilon$ for global warming  $1.5^oC$.

%\bibliographystyle{unsrt}
%\bibliography{yayunbib}

\begin{thebibliography}{10}

\bibitem{national2013abrupt}
National~Research Council,
\newblock {\em Abrupt impacts of climate change: Anticipating surprises}.
\newblock (National Academies Press, 2013).

\bibitem{NGRIP}
I.~K. Seierstad, P.~M. Abbott, M.~Bigler, T.~Blunier, A.~J. Bourne, E.~Brook,
  S.~L. Buchardt, C.~Buizert, H.~Clausen, E.~Cook, et~al.,
\newblock Consistently dated records from the greenland $\textsc{GRIP}$,
  $\textsc{GISP2}$ and $\textsc{NGRIP}$ ice cores for the past 104 ka reveal
  regional millennial-scale $\delta^{18}o$ gradients with possible heinrich
  event imprint.
\newblock {\em Quaternary Sci. Rev.} \textbf{106}, 29--46 (2014).

\bibitem{niklas2018do}
N.~Boers, M.~Ghil, and D.~D. Rousseau,
\newblock Ocean circulation, ice shelf, and sea ice interactions explain
  \textsc{D}ansgaard--\textsc{O}eschger cycles.
\newblock {\em PNAS} \textbf{115}, E11005--E11014 (2018).

\bibitem{ashwin2012tipping}
P.~Ashwin, S.~Wieczorek, R.~Vitolo, and P.~Cox,
\newblock Tipping points in open systems: bifurcation, noise-induced and
  rate-dependent examples in the climate system.
\newblock {\em Philos. T. Roy. Soc. A} \textbf{370}, 1166--1184 (2012).

\bibitem{lenton2008tipping}
T.~M. Lenton, H.~Held, E.~Kriegler, J.~W. Hall, W.~Lucht, S.~Rahmstorf, and
  H.~J. Schellnhuber,
\newblock Tipping elements in the earth's climate system.
\newblock {\em PNAS} \textbf{105}, 1786--1793 (2008).

\bibitem{ditlevsen1999}
P.~D. Ditlevsen,
\newblock Observation of $\alpha$-stable noise induced millennial climate
  changes from an ice-core record.
\newblock {\em Geophys. Res. Lett.} \textbf{26}, 1441--1444 (1999).

\bibitem{dakos2008slowing}
V.~Dakos, M.~Scheffer, E.~van Nes, V.~Brovkin, V.~Petoukhov, and H.~Held,
\newblock Slowing down as an early warning signal for abrupt climate change.
\newblock {\em PNAS} \textbf{105}, 14308--14312 (2008).

\bibitem{scheffer2009early}
M.~Scheffer, J.~Bascompte, W.~A. Brock, V.~Brovkin, S.~R. Carpenter, V.~Dakos,
  H.~Held, E.~H. Van~Nes, M.~Rietkerk, and G.~Sugihara,
\newblock Early-warning signals for critical transitions.
\newblock {\em Nature} \textbf{461}, 53 (2009).

\bibitem{lenton2011early}
T.~M. Lenton,
\newblock Early warning of climate tipping points.
\newblock {\em Nat. Clim. Change} \textbf{1}, 201 (2011).

\bibitem{FW1998}
M.~I. Freidlin and A.~D. Wentzell,
\newblock Random perturbations.
\newblock In {\em Random perturbations of dynamical systems}. (Springer, 1998).

\bibitem{wan2013hybrid}
X.~Wan and G.~Lin,
\newblock Hybrid parallel computing of minimum action method.
\newblock {\em Parallel Comp.} \textbf{39}, 638--651 (2013).

\bibitem{hunt1981path}
K.~L.~C. Hunt and J.~Ross,
\newblock Path integral solutions of stochastic equations for nonlinear
  irreversible processes: the uniqueness of the thermodynamic
  \textsc{L}agrangian.
\newblock {\em J. Chem. Phys.} \textbf{75}, 976--984 (1981).

\bibitem{durr1978om}
D.~D{\"u}rr and A.~Bach,
\newblock The \textsc{O}nsager-\textsc{M}achlup function as lagrangian for the
  most probable path of a diffusion process.
\newblock {\em Commun. Math. Phys.} \textbf{60}, 153--170 (1978).

\bibitem{onsager1953fluctuations}
L.~Onsager and S.~Machlup,
\newblock Fluctuations and irreversible processes.
\newblock {\em Phys. Rev.} \textbf{91}, 1505 (1953).

\bibitem{berkov1998mep}
D.~V. Berkov,
\newblock Numerical calculation of the energy barrier distribution in
  disordered many-particle systems: the path integral method.
\newblock {\em J. Magn. Magn. Mater.} \textbf{186}, 199--213 (1998).

\bibitem{corral2019power}
{\'A}.~Corral and {\'A}.~Gonz{\'a}lez,
\newblock Power law size distributions in geoscience revisited.
\newblock {\em Earth Space Sci.} \textbf{6}, 673--697 (2019).

\bibitem{sato1999levy}
K.~I. Sato,
\newblock {\em \textsc{L}{\'e}vy processes and infinitely divisible
  distributions}.
\newblock (Cambridge University Press, 1999.)

\bibitem{chao2018onsager}
Y.~Chao and J.~Duan,
\newblock The \textsc{O}nsager-\textsc{M}achlup function as lagrangian for the
  most probable path of a jump-diffusion process.
\newblock {\em Nonlinearity} (2019).

\bibitem{gao2016fokker}
T.~Gao, J.~Duan, and X.~Li,
\newblock Fokker-planck equations for stochastic dynamical systems with
  symmetric \textsc{L}{\'e}vy motions.
\newblock {\em Appl. Math. and Comput.} \textbf{278}, 1--20 (2016).

\bibitem{sun2017marcus}
X.~Sun, X.~Li, and Y.~Zheng,
\newblock Governing equations for probability densities of \textsc{M}arcus
  stochastic differential equations with \textsc{L}{\'e}vy noise.
\newblock {\em Stoch. Dynam.} \textbf{17}, 1750033 (2017).

\bibitem{gao2016zakai}
T.~Gao, J.~Duan, X.~Kan, and Z.~Cheng,
\newblock Dynamical inference for transitions in stochastic systems with
  $\alpha$-stable \textsc{L}{\'e}vy noise.
\newblock {\em J. Phys. A-Math. Theor.} \textbf{49}, 294002 (2016).

\bibitem{cheng2016mp}
Z.~Cheng, J.~Duan, and L.~Wang,
\newblock Most probable dynamics of some nonlinear systems under noisy
  fluctuations.
\newblock {\em Commun. Nonlinear Sci.} \textbf{30}, 108--114 (2016).

\bibitem{wang2018bifur}
H.~Wang, X.~Chen, and J.~Duan,
\newblock A stochastic pitchfork bifurcation in most probable phase portraits.
\newblock {\em Int. J. Bifurcat. Chaos} \textbf{28}, 1850017 (2018).

\bibitem{zheng2016delay}
Y.~Zheng and X.~Sun,
\newblock Governing equations for probability densities of stochastic
  differential equations with discrete time delays.
\newblock {\em Discrete Cont. Dyn-B} \textbf{22}, 3615 (2017).

\bibitem{saltzman2001dynamical}
B.~Saltzman,
\newblock {\em Dynamical paleoclimatology: generalized theory of global climate
  change}.
\newblock (Elsevier, 2001).

\bibitem{kaper2013matclimat}
H.~Kaper and H.~Engler,
\newblock {\em Mathematics and climate}.
\newblock (SIAM, 2013).

\bibitem{hoffman1998neoproterozoic}
P.~F. Hoffman, A.~J. Kaufman, G.~P. Halverson, and D.~P. Schrag,
\newblock A neoproterozoic snowball earth.
\newblock {\em Science}  \textbf{281}, 1342--1346 (1998).

\bibitem{imkeller2001ebm}
P.~Imkeller,
\newblock Energy balance models-viewed from stochastic dynamics.
\newblock In {\em Stochastic Climate Models} (Springer, 2001), pp. 213--240.

\bibitem{hansen2005capacity}
J.~Hansen, L.~Nazarenko, R.~Ruedy, M.~Sato, J.~Willis, A.~Del~Genio, D.~Koch,
  A.~Lacis, K.~Lo, S.~Menon, T.~Novakov, J.~Perlwitz, G.~Russell, G.~A.
  Schmidt, and N.~Tausnev,
\newblock Earth's energy imbalance: Confirmation and implications.
\newblock {\em Science} \textbf{308}, 1431--1435 (2005).

\bibitem{IPCC}
V.~Masson-Delmotte, P.~Zhai, H.~O. P\"ortner, D.~Roberts, J.~Skea, P.~R.
  Shukla, A.~Pirani, W.~Moufouma-Okia, C.~P\'ean, R.~Pidcock, S.~Connors,
  J.~P.~R. Matthews, Y.~Chen, X.~Zhou, M.~I. Gomis, E.~Lonnoy, T.~Maycock,
  M.~Tignor, and T.~Waterfield~(eds.),
\newblock {\em Global Warming of 1.5°C. An IPCC Special Report on the impacts
  of global warming of 1.5°C above pre-industrial levels and related global
  greenhouse gas emission pathways, in the context of strengthening the global
  response to the threat of climate change, sustainable development, and
  efforts to eradicate poverty}.
\newblock (World Meteorological Organization, Geneva, Switzerland, 2018).

\bibitem{huang2019characterization}
Y.~Huang, Y.~Chao, S.~Yuan, and J.~Duan,
\newblock Characterization of the most probable transition paths of stochastic
  dynamical systems with stable \textsc{L}{\'e}vy noise.
\newblock {\em J. Stat. Mech-Theory E.} \textbf{6}, 063204 (2019).

\bibitem{klebaner2012introduction}
F.~C. Klebaner,
\newblock {\em Introduction to stochastic calculus with applications}.
\newblock (World Scientific Publishing Company, 2012).

\bibitem{duan2015introduction}
J.~Duan,
\newblock {\em An introduction to stochastic dynamics}.
\newblock (Cambridge University Press, 2015).

\bibitem{wan2017dynamic}
X.~Wan and H.~Yu,
\newblock A dynamic-solver--consistent minimum action method: With an
  application to 2\textsc{D} \textsc{N}avier-\textsc{S}tokes equations.
\newblock {\em J. Comput. Phys.} \textbf{331}, 209--226 (2017).

\bibitem{zheng2016genetic}
Y.~Zheng, L.~Serdukova, J.~Duan, and J.~Kurths,
\newblock Transitions in a genetic transcriptional regulatory system under
  \textsc{L}{\'e}vy motion.
\newblock {\em Sci. Rep.} \textbf{6}, 29274 (2016).

\bibitem{nordhaus1992optimal}
W.~D. Nordhaus,
\newblock An optimal transition path for controlling greenhouse gases.
\newblock {\em Science} \textbf{258}, 1315--1319 (1992).

\bibitem{hasselmann1976stochastic}
K.~Hasselmann,
\newblock Stochastic climate models part $\textsc{I}$. theory.
\newblock {\em Tellus} \textbf{28}, 473--485 (1976).

\bibitem{ditlevsen1999anomalous}
P.~D. Ditlevsen,
\newblock Anomalous jumping in a double-well potential.
\newblock {\em Phys. Rev. E} \textbf{60}, 172 (1999).

\bibitem{applebaum2009levy}
D.~Applebaum,
\newblock {\em L{\'e}vy processes and stochastic calculus}.
\newblock (Cambridge University Press, 2009).

\end{thebibliography}

\section*{Acknowledgements}
We would like to thank Xiaoli Chen, Xiujun Cheng and Yang Liu for discussions about computation. This work was supported by the  NSFC grants  11801192, 1620449, Leibniz-DAAD research fellowships 2018(57423756) and Hubei provincial postdoctoral science and technology activity project.

\textbf{Author Contributions:} Y. Zheng and J. Duan designed the research and wrote the first draft of the manuscript. Y. Zheng, F.Yang and L. Fu performed computations. X.Sun and J.Kurths analysed the results and concepts development. All authors conducted research discussions and reviewed the manuscript.

\textbf{Competing interests: }The authors declare no competing financial interests.

\section*{Supplementary Materials}
\label{SI}

\renewcommand{\theequation}{SM.\arabic{equation}}
\renewcommand{\thefigure}{S.\arabic{figure}}
\setcounter{equation}{0}
\setcounter{figure}{0}

\medskip
\noindent \textbf{S1.  The    derivation of  the conditional probability density in  Eq.(\ref{p3})}

We assume that the SDE (\ref{p0}) has a unique strong solution, and the probability density for this solution exists and is strictly positive, then the conditional density of $X(t)$ given for both values of $X(0)$ and $X(t_f)$ exists. In fact, by the Markov property of SDE (\ref{p0}), the density is exactly the same as the density of $X(t_f)$ under the condition that only the value of $X(t)$ is given, i.e.,
\begin{align}\label{p4}
p(X(t_f)=x_{f} | X(0)=x_0; X(t)=x)  &= p(X(t_f)=x_{f} |  X(t)=x)\notag\\
&= Q (x_{f}, t_f |  x, t).
\end{align}
The conditional density of $X(t)$ is given by
\begin{align}\label{p5}
  p(X(t)=x | X(0)=x_0; X(t_f)=x_{f}) &=  \frac{p(X(t)=x ; X(0)=x_0; X(t_f)=x_{f}) }{p(X(0)=x_0; X(t_f)=x_{f}) }\notag\\
  &=  \frac{p(X(t_f)=x_{f} | X(0)=x_0; X(t)=x) p(X(t)=x | X(0)=x_0) }{p(X(t_f)=x_{f} | X(0)=x_0) }.
 \end{align}
Eq.(\ref{p5}) indicates that the density for $ X(t)$ defined by SDE (\ref{p0}) exists with respect to the condition $X(0)=x_0$ and $X(t_f)=x_{f}$ ,  and the right hand side of Eq(\ref{p5}) is well defined by the assumption.
 Substituting  Eq.(\ref{p4}) into Eq.(\ref{p5}), and change the notation $p$ to $Q$, we obtain the  expression for the probability density function $\mathcal{P}_{A}(x, t)$
 \begin{align}\label{p6}
\mathcal{P}_{A}(x, t) &= p(X(t)=x | X(0)=x_0; X(t_f)=x_{f}) \notag\\
&=  \frac{Q(x_{f}, t_f | x, t)Q(x, t | x_0, 0)}{Q(x_{f}, t_f | x_0, 0)} .
 \end{align}
Thus, a governing equation for the transition probability density $\mathcal{P}_{A}(x, t)$ of the solution of  SDEs(1)   is derived. See Fig.\ref{fig:1.2}(A).

 \medskip
\noindent \textbf{S2.  A simple example with    analytical solution for the maximum likelihood  transition path.}

 In order to verify that the numerical scheme is valid,  the numerical solution is compared with the analytical solution for the  maximum likelihood  transition path  $x_m(t)$. We consider the following scalar SDE  with additive  Gaussian noise
\begin{equation} \label{ou}
{\rm d}X(t) = -aX(t){\rm d}t+ b{\rm d}B_t ;  \quad X(0)=x_0 \in \mathbb{R}^1.
\end{equation}
where $a, b$ are real parameters. In the special case of (\ref{ou}) for $a = 1$, $b = 0.1$, $x_0 = -1$, $x_f=1$, $T_f=10$ and a bounded domain $D=(-5, 5)$. Based on the expression of the conditional  probability density $\mathcal{P}_{A}(x, t)$  in  Eq.(\ref{p3}), we may calculate the
the density function  $Q(x, t | x_0, 0)$ of SDE(\ref{ou})
\begin{align*}
Q(x, t | x_0, 0)
&= \frac{\sqrt{a}}{\sqrt{\pi b^2(1-e^{-2at})}} \exp[-\frac{a(x-e^{-at}x_0)^2}{b^2(1-e^{-2at})}],
\end{align*}
and the other density function $Q(x_{f}, t_f | x, t)$ is given by
\begin{align*}
Q(x_{f}, t_f | x, t) &= \frac{\sqrt{a}}{\sqrt{\pi b^2(1-e^{-2a(t_f-t)})}} \exp[-\frac{a(x_{f}-e^{-a(t_f-t)}x)^2}{b^2(1-e^{-2a(t_f-t)})}],
\end{align*}
since the production of  the density function $Q(x_{f}, t_f | x, t)Q(x, t | x_0, 0)$ is a strictly increasing function with $x$, thus the maximum exists in a bounded domain.  The derivatives of the density function with respect to $x$, i.e., the analytical solution,  is solved
\begin{align*}
x_{max}(t)
&= \frac{(e^{-a(t_f-t)} - e^{-a(t_f+t)})x_{f}+ (e^{-at} - e^{-a(2t_f-t)})x_0}{1-e^{-2at_f}}.
\end{align*}
The numerical solution $x_{max}(t)$ can be simulated via numerical global optimization of $\mathcal{P}_{A}(x, t)$ (Eq.(\ref{p3})) by numerical algorithm of Gao \emph{et al.}\cite{gao2016fokker}.  See Fig.\ref{fig:1.2}(B).

\medskip
\noindent\textbf{S3. The $\alpha$-stable  L\'evy process.}
The well-known Brownian motion is a Gaussian process, with stationary and independent increments, and almost surely continuous sample paths. 
A  L\'evy process $L_t$  is a non-Gaussian  process, with stationary and independent increments, i.e., for any $s, t$ with $0\leq s \leq t$, the distribution of $L(t) - L(s)$ only depends on $t - s$, and for any partition $0=t_{0} < t_{1} < \cdots < t_{n}=t$, $L(t_{i}) - L(t_{i-1})$, $i=1,2,\cdots,n$ are independent. The sample paths of Levy process are almost surely right continuous with left limits ($c\grave{a}dl\grave{a}g$)\cite{duan2015introduction}, and as a result, the sample paths have countable jumps.  L\'evy processes are thought as    appropriate models for non-Gaussian fluctuations with heavy tailed statistical distributions and intermittent bursts. The characteristic function for a L\'{e}vy process in   $\mathbb{R}^d$  with a generating triplet $(b, Q, \nu)$ is given by the L\'{e}vy - Khintchine formula \cite{applebaum2009levy},
\begin{equation*}
Ee^{i <\lambda, L_t>} = \exp \Big\{ it <b, \lambda>  - t\frac{1}{2} <\lambda, Q\lambda>+ t\int_{\mathbb{R}^{d}\backslash \{0\}}(e^{i<\lambda, y>} - 1 - i<\lambda, y> I_{\|y\|<1} y) \nu(dy)  \Big\},
\end{equation*}
where the notation $<\cdot , \cdot>$ is the inner product in $\mathbb{R}^d$, the $I_S$ is the indicator function of the set $S$. Thus,   the L\'{e}vy process  is characterized by a vector  $b \in \mathbb{R}^{d}$, a positive definite symmetric $d \times d$ matrix $Q$ and a L\'{e}vy  jump measure $\nu$ defined on $\mathbb{R}^{d}\backslash\{0\}$  satisfying
\begin{equation}
  \int_{\mathbb{R}^{d}\backslash \{0\}}(\|y\|^{2} \wedge 1) \nu(dy) < \infty. \notag
\end{equation}
The L\'{e}vy  jump measure quantifies the jump frequency and size for sample paths of this  L\'evy process.

A $\alpha$-stable  L\'evy process $L^\alpha_t$ is a special type of L\'evy process defined by the stable L\'evy random variable with the distribution $S_\alpha(\delta, \beta, \lambda)$. Usually, $\alpha \in (0, 2)$ is called the L\'evy index (non-Gaussianity  index), $\delta \in [0, \infty)$ is the scale parameter, $\beta \in [-1, 1]$ is the skewness parameter and $\lambda \in (-\infty, \infty)$ is the shift parameter.

 A stable L\'evy random variable $L^\alpha$  has the following  ``\emph{heavy tail} '' estimate
\begin{equation*}
\lim_{y \rightarrow \infty}y^{\alpha}\mathbb{P}(L^\alpha > y) = C_{\alpha}\frac{1 + \beta}{2}\sigma^{\alpha},
\end{equation*}
i.e., the tail estimate decays in a power law. The constant  $C_{\alpha}$ depends on $\alpha$.

In particular, for a symmetric $\alpha$-stable L\'{e}vy process $L^\alpha_t$ with the generating triplet  $(0, 0,  \nu_{\alpha})$, the characteristic function  becomes
\begin{equation*}
Ee^{i<\lambda, L^\alpha_t>} = e^{-t\|\lambda\|^\alpha}, \quad t > 0, \lambda \in \mathbb{R}^d,
\end{equation*}
with  the jump measure \cite{sato1999levy}
\begin{equation*}
 \nu_{\alpha}(dy) = \frac{C(\alpha, d)}{\|y\|^{n + \alpha}}dy.
\end{equation*}
The constant  $C_{\alpha, d}$ depends on $\alpha$ and dimension $d$.
The $\alpha$-stable L\'evy motion $L^\alpha_t$  has larger jumps with lower jump probabilities for $\alpha$  is small ($ 0 < \alpha <1$), while it has smaller jumps with higher jump frequencies for large $\alpha$
 values  ($1 < \alpha <2$).
 The special case $\alpha  = 2$ corresponds to the usual Brownian motion, which is a Gaussian process.

%There are three equilibrium states are $T* \approx 288K$, $T* \approx 265K$, $T* \approx 233K$ for $\epsilon = 0.6$.

\begin{figure}[!t]
\begin{minipage}[b]{ \textwidth}
%\leftline{(a)}
\centerline{\includegraphics[height = 10cm, width = 0.9\textwidth]{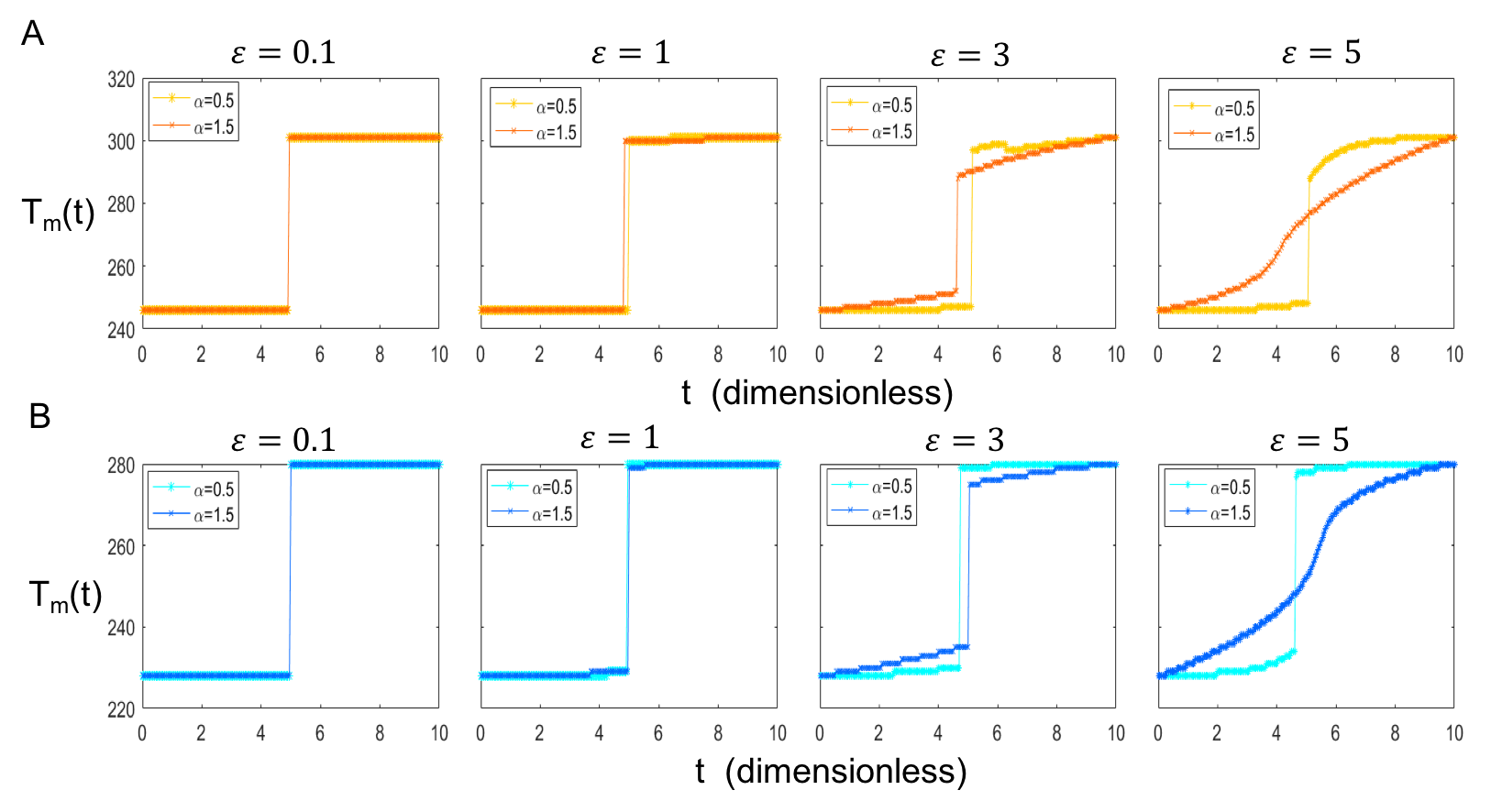}}
\end{minipage}
\caption{\textbf{The dependence of the maximum likelihood climate change on the $\alpha$-stable L\'evy noise intensities $\epsilon$ for global warming  $1.5^oC$. }The maximum likelihood  path for:  (A) $\gamma=0.51$ (transition from the cold climate stable state $T=245.7\rm K$ to the warmer one $T=301.6\rm K$) and (B) for $\gamma=0.67$ (transition from the deep-frozed climate stable state $T=228\rm K$ and the warmer one $T=279.7\rm K$) with $\alpha=0.5$ and $\alpha=1.5$.}
\end{figure}

\end{document}